# Concatenated Convolution-Polar codes over Rayleigh Channels degraded as Erasure Channels

Mohammed Usman
Department of Electrical Engineering, King Khalid University
Abha, Saudi Arabia
Jaypee University of Information Technology (JUIT)
Solan, India
omfarooq@kku.edu.sa

*Abstract*—Polar codes have been shown to approach capacity of symmetric binary erasure channels and also have low encoding and decoding complexity. Wireless channels are bursty in nature and a method to apply polar codes over wireless channels is proposed. This paper investigates the performance of polar codes over typical urban fading wireless channels. The wireless channel is modeled as a degraded erasure channel by using an inner convolution code and interleaving. When the convolution code fails to correct any errors, the corresponding bits are discarded and treated as erasures. We present the performance of polar codes over such degraded wireless channels using different code rates for the inner convolution code.

## I. INTRODUCTION

Reliability in communication is usually achieved using forward error correction (FEC), automatic retransmission request (ARQ) or a combination of both. Several FEC schemes have been developed over the years, each having their own advantages and disadvantages. A fundamental problem which coding theorists have been trying to address is to construct codes to transmit data at a rate that approaches the capacity of the channel as described in the classical paper of Shannon in 1948. Shannon showed in his paper that there exist codes using which it is possible to transmit data at a rate below the capacity $C$ of the channel (memoryless) with arbitrarily small probability of error [1]. Several different codes have been developed which narrow the gap between achievable rate and $C$ for various considerations of channel realization. A comprehensive historical development of coding theory is available in [2]. More recently, Erdal Arikan developed polar codes that approach capacity of any binary memoryless channels. Polar codes further have linear encoding and decoding complexity in the block length 'N' as $O(N \log N)$ [3]. Since fading wireless channels are channels with memory, we use an inner convolution code followed by interleaver to degrade the fading wireless channel into a simple erasure channel. The average erasure probability $\varepsilon$ is then obtained through bit level simulations for mobiles traveling at pedestrian speeds (5 km/hr) and vehicular speeds (50 km/hr). Polar codes are then used over the degraded Rayleigh channel and the trade-off between the rate of inner code and polar code is analyzed. The rest of the paper is organized as follows: Section II provides a description of polar codes along with coding theoretic definitions and polar encoding. Section III presents the method used for degrading the fading wireless channel into an erasure channel. Section IV presents a brief description of the successive cancellation (SC) decoder for polar codes. Simulation results for the degraded wireless channel are discussed in section V. Conclusions and further work are discussed in section VI.

## II. POLAR CODES

Invented in 2009, by Erdal Arikan, polar codes achieve capacity over the class of symmetric binary discrete memoryless channels (B-DMC). Further, the encoding and decoding complexity of polar codes scales linearly in the block length as $O(N \log N)$. For a given B-DMC, $W$, polar codes operate on the principle of synthesizing $N = 2^n$ different channels from a set of $N$ independent copies of $W$ through a process called channel polarization.. The $i^{th}$ channel among the $N$ different polarized channels is denoted as $W_N^{(i)}$ for $\{1 \leq i \leq N\}$. I(W) denotes the symmetric capacity of the channel W and correspondingly, $I(W_N^{(i)})$ denotes the symmetric capacity of the polarized channel $W_N^{(i)}$. Channel polarization has the effect that the capacity of each polarized channel tends towards 0 or 1 as N becomes large. Of the N polarized channels, K = N I(W) channels have capacity 1 and the remaining N – K = N(1 – I(W)) have capacity 0; i.e. there are K noiseless channels which are reliable and N – K noisy channels which are unreliable. Polar coding is implemented by transmitting K data bits over the K noiseless channels and N – K frozen bits, known to the decoder, over the N – K noisy channels [3]. The set of K reliable channels whose capacity approaches 1 is denoted as $\chi$ and the set of N - K noisy channels whose capacity approaches 0 is represented by the complementary set of $\chi$ and is denoted as $\chi^C$.

The underlying math of polar codes design opens up a new mathematical framework to address the channel coding problem, left open by Shannon, i.e. codes that achieve capacity with reasonable complexity.

The polar coding technique involves a linear transformation, called polar transform, which is applied to an N-element channel vector. The linear transformation is implemented by taking several Kronecker tensor products of a chosen matrix $G_2$ defined as [3]

$$G_2 = \begin{bmatrix} 1 & 0 \\ 1 & 1 \end{bmatrix}$$

(1)

The matrix $G_N$ is obtained by taking $n = log_2(N)$ Kronecker tensor products of $G_2$. The polar transformation is applied by multiplying the channel vector with the transformation matrix $G_N$(modulo 2), resulting in a set of polarized channels $W_N^{(i)}$, such that except for a small fraction '$i$', the mutual information of each $W_N^{(i)}$ tends to either 0 or 1 for large $N$. The mutual information of a B-DMC having input alphabet X, which is i.i.d Bernoulli distributed and output alphabet Y, is defined as [4]

$$I\left(W_N^{(i)}\right) = I(X,Y) \triangleq \frac{1}{2} \sum_{y \in Y} \sum_{x \in X} p(y|x) \log \frac{p(y|x)}{\frac{1}{2}p(y|0) + \frac{1}{2}p(y|1)}$$

(2)

with each $I(W_N^{(i)})$ approaching either 0 or 1. It must be noted that the polarization phenomenon finds application in source coding as well.

*A. Polar Encoding*

The generator matrix on an (N, K) polar code to be used over a binary erasure channel is a K x N sub-matrix of $G_N = G_2^{\otimes n} = \begin{bmatrix} 1 & 0 \\ 1 & 1 \end{bmatrix}^{\otimes n}$, where $\otimes$ denotes the Kronecker tensor product for any $N = 2^n$ with $n \geq 1$ and $1 \leq K \leq N$. The generator matrix $G_N$ is obtained by first calculating, for $k = 2^0, 2^1, 2^2......2^{n-1}$, a vector of N Bhattacharya parameters corresponding to N bit channels in a recursive manner as described in [5]. The Bhattacharya parameters in each vector of the set of N vectors are permuted to be arranged in ascending order. The generator matrix is then the sub-matrix of $G_N$, denoted as $G_N(\chi)$ containing the rows with indices corresponding to the first K elements of the permuted Bhattacharya parameter set.

The input block $\bar{u} = (u_\chi, u_{\chi C})$ consists of the K information bits $u_\chi$ and N-K frozen bits $u_{\chi C}$. The codeword is then $\bar{x} = \bar{u} G_N = u_\chi G_N(\chi) + u_{\chi C} G_N(\chi C)$. Since $u_{\chi C}$ is a set of fixed bits, $\bar{x} = u_\chi G_N(\chi) + c$ where 'c' is a fixed vector defined as $c = u_{\chi C} G_N(\chi C)$. $\bar{x}$ represents the input to the channel and the corresponding channel output is denoted as $\bar{y}$.

III. DEGRADED RAYLEIGH FADING CHANNEL

Polar codes have been shown to perform best over symmetric binary memory-less channels. Since fading wireless channels are channels with memory, an indirect method is used in this paper to degrade a fading wireless channel into an erasure channel. The erasure probability of the degraded wireless channel is estimated through link level simulations. The operating frequency is considered to be in the 1 GHz band. Considering binary channel, the modulation scheme used is Gaussian Minimum Shift Keying (GMSK). In order to degrade the fading wireless channel with memory into a memory-less erasure channel, an inner convolution code is used followed by interleaving. At the receiver side, the received bits are first de-interleaved followed by decoding of the convolution code using soft decision Viterbi decoding. A Block Check Sequence (BCS) is used to detect residual errors after decoding the convolution code. If the BCS fails, the complete data block is discarded and the corresponding bits are treated as erasures. The wireless channel model used in this work considers a mobile unit moving at pedestrian speed (5 km/hr) and vehicular speed (50 km/hr) in typical urban scenarios. Dynamic variations of the signal to interference ratio (SIR) follow a log-normal distribution using a de-correlation distance of 20 m [6].

The convolution codes are constructed with three different code rates i.e. ½, 2/3 and ¾. Details of the convolution code and interleaving are available in [7-9]. The average erasure probability of the degraded wireless channel for different code rates of the convolution code are summarized in table I.

TABLE I. ERASURE PROBABILITY OF DEGRADED RAYLEIGH CHANNEL

| Code rate / Mobile speed | 1/2 | 2/3 | 3/4 |
|---|---|---|---|
| Pedestrian | 0.054 | 0.078 | 0.093 |
| Vehicular | 0.014 | 0.035 | 0.063 |

It is noticed that the erasure probability is lower at higher mobile speed. This is attributed to the spreading of errors at higher speeds resulting in shorter bursts of errors which are corrected by the convolution code. The erasure probability obviously increases as the code rate is made larger, since a higher code rate means lower error correction capability of the convolution code.

The reliability of a channel may be specified by the Bhattacharya parameter, β, defined as [4]

$$\beta = \sum_{y \in Y} \sqrt{p(y/0)p(y/1)}$$

(3)

For a binary erasure channel with erasure probability ε, the Bhattacharya parameter β = ε. Degradation of a fading channel as an erasure channel is achieved at the cost of additional overhead and complexity associated with the convolution code. The degraded Rayleigh channel is then used to study the performance of polar codes and the results are discussed in section V.

The set of K reliable erasure channels synthesized by the degradation of the Rayleigh fading channel is specified such that $I\left(W_N^{(i)}\right) \geq I\left(W_N^{(j)}\right)$ for all $i \in \chi$ and $j \in \chi^C$. In terms of the Bhattacharya parameter, this is equivalently specified as $\beta\left(W_N^{(i)}\right) \leq \beta\left(W_N^{(j)}\right)$ for all $i \in \chi$ and $j \in \chi^C$.

## IV. POLAR DECODING

The successive cancellation (SC) decoder is a low complexity decoder for polar codes. The decoder must estimate $\hat{u}$ by observing the received vector $\bar{y}$. The estimate of each element of the transmitted block $\hat{u}_i$ for $1 \leq i \leq N$ is obtained as described in [5]. For a frozen bit $\hat{u}_i$, the decoder sets its value equal to the known value. For an information bit, the decoder estimates all the previous bits and computes a likelihood ratio, sets the information bit equal to 0 if the likelihood ratio is greater than 1 and sets the information bit equal to 1, otherwise.

$$\hat{u}_i = \begin{cases} u_i & if\ i \in \chi^c \\ h_i(y_1^N, \hat{u}_1^{i-1}) & if\ i \in \chi \end{cases} \quad (4)$$

where

$$h_i(y_1^N, \hat{u}_1^{i-1}) \triangleq \begin{cases} 0 & if\ \dfrac{W_N^{(i)}(y_1^N, \hat{u}_1^{i-1}|0)}{W_N^{(i)}(y_1^N, \hat{u}_1^{i-1}|1)} \geq 1 \\ 1 & otherwise \end{cases} \quad (5)$$

If the estimated data block $\hat{u}_\chi$ is different from the transmitted data block $u_\chi$, it results in a block error. An analysis of error bounds of SC decoder for various channels is presented in [10].

## V. RESULTS

In this paper, a study of the variation of code rate of polar codes to achieve a block error rate below a specified upper limit is performed, for Rayleigh channels which have been degraded as erasure channels using an inner convolution code. The price paid is the additional complexity associated with the encoding and decoding of convolution codes as well as the associated overhead data. The variation of code rate for polar code with erasure probability is shown in figures 1-4 for various block lengths (N = $2^4$, $2^8$, $2^{10}$ and $2^{12}$) and target BLER of 0.1, 0.3 and 0.5.

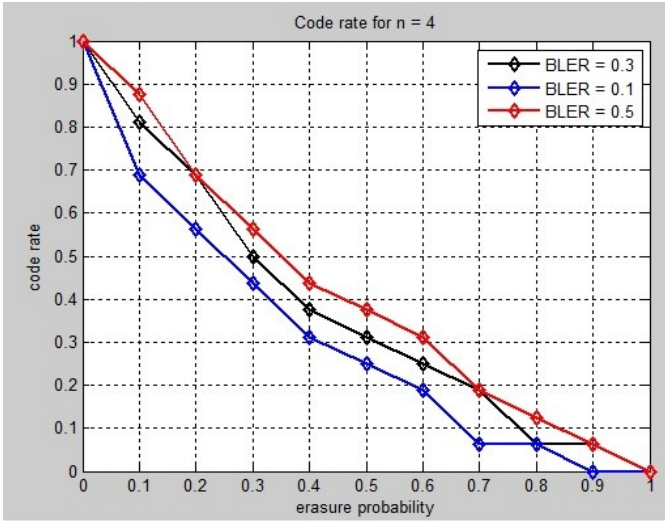

Fig. 1. Code rate for target BLER with fixed block length N = $2^4$

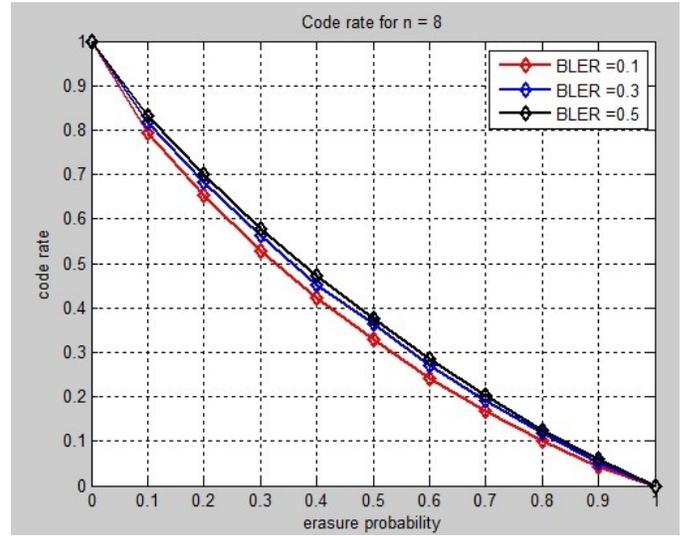

Fig. 2. Code rate for target BLER with fixed block length N = $2^8$

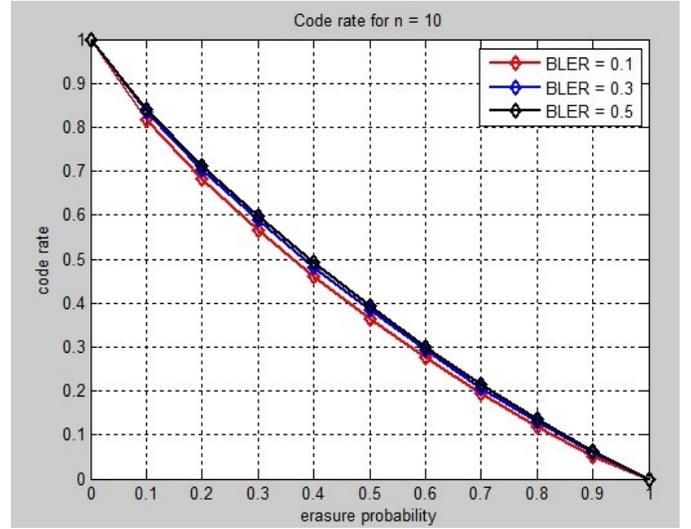

Fig. 3. Code rate for target BLER with fixed block length N = $2^{10}$

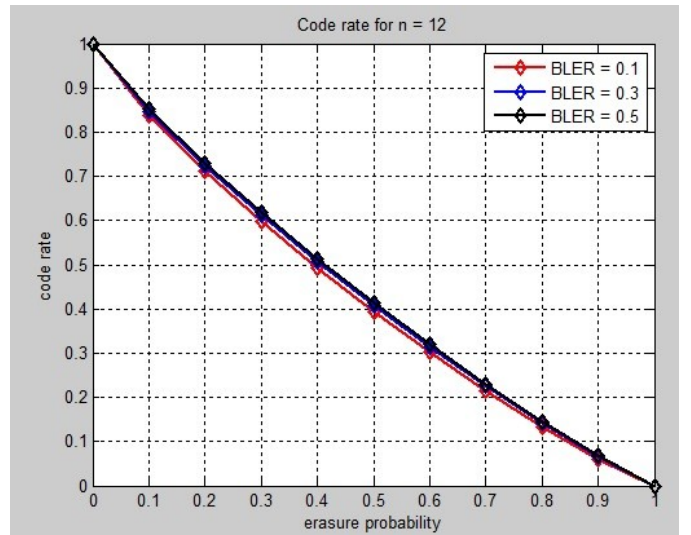

Fig. 4. Code rate for target BLER with fixed block length N = $2^{12}$

It is clear that to maintain the BLER below the specified upper limit, the polar code rate has to be lowered as the erasure probability increases. It is also noted by comparing figures 1-4, that the difference between the achievable code rate for different target BLER values becomes marginal as the block length N is increased. This means that in order to maintain the BLER within the specified target, while keeping the polar code rate high, the block length N has to be increased.

Since the encoding and decoding complexity of polar codes scales as $O(N \log N)$, larger value of N translates to increased computational complexity. By choosing a large value for block length, the target BLER performance of polar codes can be achieved while maintaining a relatively higher code rate as compared to that of a smaller block length. For purposes of clarity, the variation of polar code rate with erasure probability to achieve a target BLER of 0.1 is plotted in figure 5 for various block lengths.

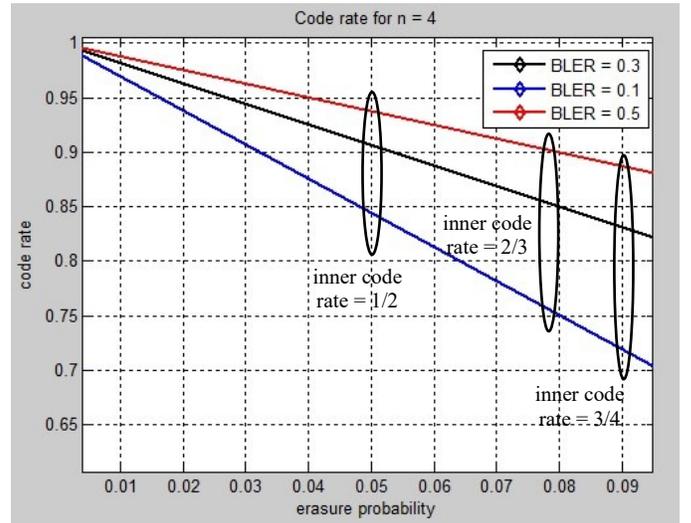

Fig. 6. Trade-off between inner code rate and polar code rate (mobile unit moving at 5 km/hour)

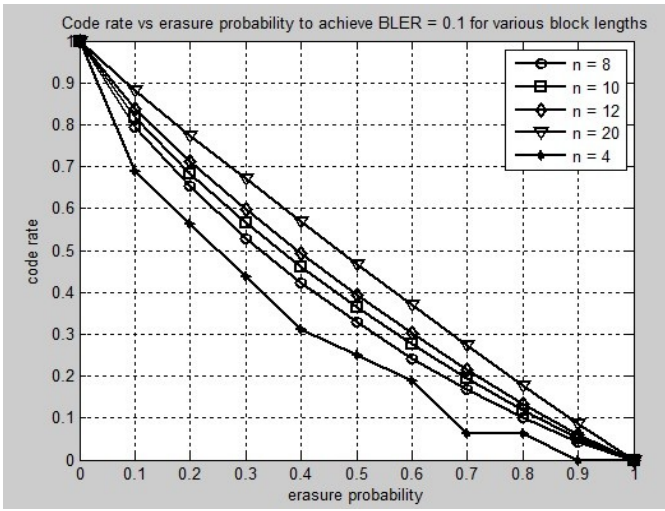

Fig. 5. Comparison of polar code rate for various block lengths $N = 2^n$

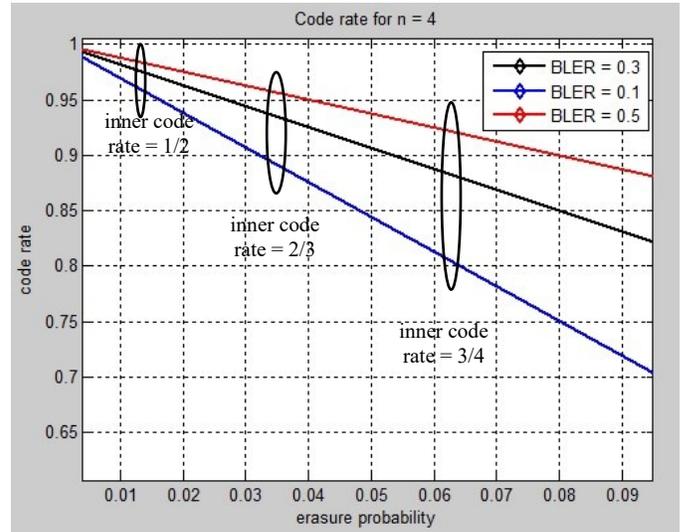

Fig. 7. Trade-off between inner code rate and polar code rate (mobile unit moving at 50 km/hour)

If the code rate of the inner code is higher, the added redundancy due to the inner code will be lower. However, higher code rate of inner code results in higher erasure probability of the degraded Rayleigh channel which in turn would require the code rate of polar codes to be lowered, meaning more overheads associated with the polar code. Hence, there is a trade-off between the code rate of the inner code and the code rate of polar code as illustrated in figures 6 and 7 for pedestrian and vehicular speeds respectively.

It is observed from figures 6 and 7 that an increase in the rate of the inner convolution code necessitates a decrease in rate of the polar code to maintain the target BLER. The actual values of the rate of inner code and the corresponding rate of polar code are given in table II for $N = 2^4$. The smallest value of N used in this study has been considered to study the trade-off between inner code rate and polar code rate in order to depict the worst case scenario.

When the reliability requirement of the system is kept high i.e. smaller target BLER, the slope for the decay of polar code rate with erasure probability is higher as compared to that of larger target BLER, as evident from figures 6 and 7. This means that higher accuracy translates to higher overheads. Hence, there is an inherent cost, in terms of overheads, associated with the demand for accuracy.

### A. Trade-off Ratio (τ)

It is noted from table 2 that the trade-off between the rate of inner code and the rate of polar code is not proportionate. For example, for a mobile moving at 5 km/hour, with a target BLER set to 0.1, an improvement of the inner code rate by about 16% causes the polar code rate to be lowered by 8%. Similarly, improving the inner code rate by 8% causes the polar code rate to be lowered by 4%. We introduce the term trade-off ratio '$\tau$' which is defined as the ratio of increase in the rate of inner code to the corresponding decrease in the rate of polar code to achieve a predefined target BLER. Thus the trade-off ratio for target BLER = 0.1 is $\tau$ = 2:1. Similarly, the trade-off ratio for all tested scenarios corresponding to pedestrian speed are summarized in table III.

TABLE II. INNER CODE RATE VS POLAR CODE RATE

| Mobile speed | Target BLER | Inner code rate | Polar code rate |
|---|---|---|---|
| 5 km/hour | 0.1 | 0.5 | 0.84 |
| | | 0.667 | 0.76 |
| | | 0.75 | 0.72 |
| | 0.3 | 0.5 | 0.905 |
| | | 0.667 | 0.851 |
| | | 0.75 | 0.828 |
| | 0.5 | 0.5 | 0.935 |
| | | 0.667 | 0.9 |
| | | 0.75 | 0.88 |
| 50 km/hour | 0.1 | 0.5 | 0.962 |
| | | 0.667 | 0.88 |
| | | 0.75 | 0.804 |
| | 0.3 | 0.5 | 0.975 |
| | | 0.667 | 0.93 |
| | | 0.75 | 0.88 |
| | 0.5 | 0.5 | 0.98 |
| | | 0.667 | 0.955 |
| | | 0.75 | 0.925 |

TABLE III. TRADE OFF RATIO ($\tau$) FOR MOBILE UNIT TRAVELING AT 5 KM/HOUR

| Mobile speed | Target BLER | Trade-off ratio '$\tau$' |
|---|---|---|
| 5 km/hour | 0.1 | 2:1 |
| | 0.3 | 3:1 |
| | 0.5 | 4:1 |

Such a pattern is not observed for a mobile unit traveling at 50 km/hour. The degradation of polar code rate is around 8% when the rate of inner code is increased from 0.5 to 0.667 and also when increased from 0.667 to 0.75, when the target BLER is 0.1. Similarly, when the target BLER is 0.3, the reduction in polar code rate is around 5%. The corresponding reduction in polar code rate with target BLER set to 0.5 is around 3%. For the chosen rates of inner code, the erasure probabilities of the corresponding degraded Rayleigh channels are more or less uniformly spaced, when the mobile unit is traveling at 50 km/hour, as observed from figure 7. However, the same is not true in the case of a mobile unit traveling at 5 km/hour as observed from figure 6. This explains why the degradation of polar code rate remains same when the rate of inner code is changed from 0.5 to 0.667 and from 0.667 to 0.75, for each of the target BLER's. In general, the results show that the loss attributed to the degradation of the rate of polar code is small compared to the gain achieved by increasing the rate of the inner convolution code. It would therefore make practical sense to use an inner code with high code rate.

VI. CONCLUSIONS AND FURTHER WORK

This paper presents a mechanism to degrade a Rayleigh fading channel and model it as an erasure channel. The channel degradation is performed by using an inner convolution code along with interleaving. The trade-off between the code rate of the inner convolution code and the outer polar code is investigated to achieve performance within various target BLER's for different block lengths of the polar code as well as for different speeds at which the mobile unit is traveling. While there is an inherent trade-off between the inner code rate and polar code rate, a detailed analysis of the results indicate that it is more efficient to employ a high rate for the inner convolution code. It is evident from the results that the reduction in the rate of polar code is much smaller for a given increase in the rate of the inner code.

Further work in the immediate future work shall focus on other frequency bands of interest such as around 2 GHz and 2.4 GHz and later on in mmWave frequency bands. The trade-off between the encoding and decoding complexities of the inner code and outer code shall also be investigated in future studies.